\documentclass[showkeys,twocolumn]{revtex4}
\usepackage{epsfig}
\usepackage{graphicx}
\begin{document}
\title[Information flow in dominance networks of social insects]{Social insect colony as a biological regulatory system: Information flow in dominance networks} 
\author{Anjan K. Nandi$^{1,}$\footnote{Corresponding author:\\Email: anjanandi@gmail.com}, Annagiri Sumana$^{2}$ and Kunal Bhattacharya$^{3}$}
\affiliation{
$^{1}$ Centre for Ecological Sciences, Indian Institute of Science, Bangalore 560012, India.\\
$^{2}$ Department of Biological Sciences, Indian Institute of Science Education and Research - Kolkata, Mohanpur 741252, India.\\
$^{3}$ Department of Physics, Birla Institute of Technology and Science, Pilani 333031, India.}
\begin{abstract}
Social insects provide an excellent platform to investigate flow of information in regulatory systems since their successful social organization is essentially achieved by effective information transfer through complex connectivity patterns among the colony members. Network representation of such behavioural interactions offers a powerful tool for structural as well as dynamical analysis of the underlying regulatory systems. In this paper, we focus on the dominance interaction networks in the tropical social wasp \textit{Ropalidia marginata} - a species where behavioural observations indicate that such interactions are principally responsible for the transfer of information between individuals about their colony needs, resulting in a regulation of their own activities. Our research reveals that the dominance networks of \textit{R. marginata} are structurally similar to a class of naturally evolved information processing networks, a fact confirmed also by the predominance of a specific substructure - the `feed-forward loop' - a key functional component in many other information transfer networks. The dynamical analysis through Boolean modeling confirms that the networks are sufficiently stable under small fluctuations and yet capable of more efficient information transfer compared to their randomized counterparts. Our results suggest the involvement of a common structural design principle in different biological regulatory systems and a possible similarity with respect to the effect of selection on the organization levels of such systems. The findings are also consistent with the hypothesis that dominance behaviour has been shaped by natural selection to co-opt the information transfer process in such social insect species, in addition to its primal function of mediation of reproductive competition in the colony.
\end{abstract}
\maketitle

\section{Introduction} 
\subsection{Information flow in biological systems}
\begin{figure*}
\centerline{\epsfig{file=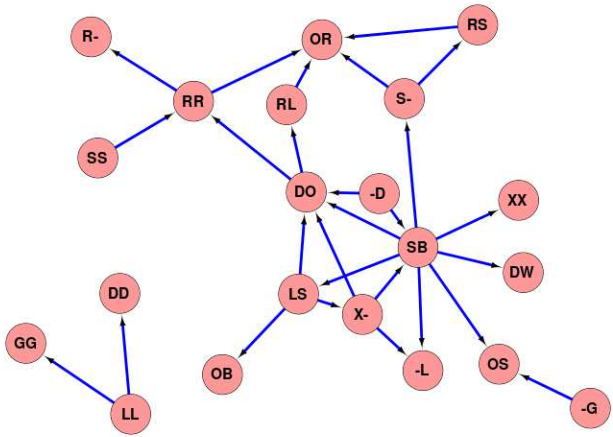,width=12.5cm}}
\caption{The dominance network in the colony V215 of \textit{R. marginata}. Nodes represent individuals with their unique identification codes, links represent dominance relationships directed from the dominant to the subordinate individual. The figure has been drawn by using Cytoscape \cite{Smoot.etal.11}.}
\end{figure*}

Living organisms are characterized by various sequential processes operating at different biological levels, such as genes, proteins, cells, neurons etc. Their survival depends heavily on the proper functioning of such coordinated processes and hence on an efficient dissemination of information through the communication systems of the respective levels \cite{Palsson.06}. Also for the human engineered systems like electronic circuits or the Internet, the primary task is to pass information from one part of the system to another \cite{Newman.10}. All of these systems are known to maximize their performance under time or energy constraints. While the structural stability and economics are responsible for the optimization of artificial systems, the biological systems are generally optimized under natural selection. Therefore, it is of particular interest to know how such systems achieve their effective process of information transfer and what factors are responsible for their efficiency.

Flow of information is an important criterion also for the coordinated activities of group living animals. Information transfer is crucial in processes like collective motion of fish schools, flocks of birds, herds of quadrupeds etc., where a single animal is inept to communicate at any moment with all the other members in a group \cite{Vicsek.Zafeiris.12}. In the human society too, processes like reaching a consensus or the spreading of rumors and diseases are governed by an effective flow of information \cite{Castel.etal.09}. Social insect species such as ants, bees, and wasps exhibit superlative forms of regulation in their colony life that rely largely on the exchange of information among the individuals \cite{Wilson.71, Holl.Wilson.09}. In such species, only one or a small subset of individuals are fertile, known as the queens, who use `honest signals' to maintain their reproductive monopoly over their workers \cite{Keller.Nonacs.93}. Some of the social insect species are more advanced and are known as `highly eusocial' species. They build large colonies and have evolved to use advanced methods of information transfer that can cater to thousands of workers. For example, trail-forming ants use chemical signals to communicate information about the foraging sites while honeybees use a dance language for the same purpose \cite{Wilson.71}. In the less advanced `primitively eusocial' species, colony sizes are small and direct physical interactions play a more significant role in the flow of information. Since, in the course of evolution, these primitively eusocial species are considered to be intermediate between their solitary and highly eusocial counterparts \cite{Wilson.71}, study of such species could shed light on the evolutionary processes by which their regulatory systems have been optimized.

\subsection{The model system}
\textit{Ropalidia marginata} is a primitively eusocial wasp widely found in the peninsular India and other south-east Asian countries \cite{Gadagkar.01}. They use several forms of pair-wise physical interactions like dominance, antennation, allogrooming etc., among which the dominance behaviour plays a major role in regulating the activities of the workers \cite{Chandra.Gadagkar.91}. Some workers are found to be specialized to perform intranidal tasks such as building of cells or brood care, while the others opt for extranidal tasks like foraging for food and building materials \cite{Gadagkar.Joshi.83}. Workers who spend most of the time on the nest to perform intranidal duties may get information about the colony needs directly by inspection and convey the same to the foragers. It has been experimentally demonstrated that the foragers receive more dominance than the non-foragers and the frequency of the dominance received by the foragers is correlated with their foraging rates \cite{Premnath.etal.95}. It was also found that the dominance received by the foragers increases when a colony is forced to starve \cite{Lamba.etal.07}, and decreases when a colony is fed in excess \cite{Bruyndonckx.etal.06}. Similar correlations between dominance behaviour and worker activities are also observed in other species of the same subfamily Polistinae: \textit{Polistes dominulus} \cite{Therau.etal.90, Therau.etal.92}, \textit{Polybia occidentalis} \cite{Odonnell.01}, and \textit{Polybia aequatorialis} \cite{Odonnell.03}. These observations lead to the hypothesis that the social wasp workers use dominance behaviour to transfer information about the colony needs among themselves and hence achieve an effective regulation of their own activities. In the present study, we investigate this hypothesis with respect to \textit{R. marginata} by analyzing the structure of the networks constructed from their interaction relationships.

\begin{table*}
\begin{tabular}{|c|c|c|c|c|c|c|c|}
\hline 
\scriptsize{Colony} & \scriptsize{Number} & \scriptsize{Number} & \scriptsize{Average} & \scriptsize{Clustering} & \scriptsize{Assortativity} & \multicolumn{2}{|c|}{\scriptsize{$P$ for $\chi^2$ goodness of}}\\ 
\scriptsize{identity} & \scriptsize{of} & \scriptsize{of} & \scriptsize{path} & \scriptsize{coefficient} & \scriptsize{coefficient} & \multicolumn{2}{|c|}{\scriptsize{fit to Poisson distribution}}\\ \cline{7-8}
  & \scriptsize{nodes} & \scriptsize{links} & \scriptsize{length} &  &  & \scriptsize{In-degree} & \scriptsize{Out-degree}\\
 & $(N)$ & $(L)$ & $(\bar{l})$ & $(C)$ & $(r)$ & \scriptsize{distribution} & \scriptsize{distribution}\\
\hline 
V213 & 43 & 64 & 3.66 & 0.07 & -0.04 & 0.22 & 0.42\\ 
\hline
V215 & 21 & 27 & 1.88 & 0.12 & -0.25 & 0.26 & 0.18\\ 
\hline 
V217 & 29 & 32 & 2.25 & 0.00 & -0.14 & 0.15 & 0.41\\ 
\hline 
V219 & 14 & 22 & 2.10 & 0.20 & -0.29 & 0.98 & 0.76\\ 
\hline 
V220 & 15 & 19 & 1.21 & 0.10 & -0.14 & 0.67 & 0.90\\ 
\hline 
V221 & 14 & 18 & 1.37 & 0.14 & -0.51 & 0.03 & 0.55\\ 
\hline 
V222 & 20 & 41 & 1.85 & 0.21 & -0.37 & 0.83 & 0.09\\ 
\hline 
V223 & 17 & 26 & 1.85 & 0.16 & -0.08 & 0.53 & 0.35\\ 
\hline 
V224 & 14 & 23 & 1.68 & 0.20 & -0.21 & 0.62 & 0.48\\ 
\hline 
\end{tabular} 
\caption{Basic global structural quantities measured on the dominance networks of \textit{R. marginata}.}
\end{table*}

\subsection{Network substructures}
Over the past two decades, network science has emerged as a very powerful tool with regard to the analysis of complex systems \cite{Newman.10}. Study of different biological and other real-world networks have revealed that many of the networks share global statistical features such as existence of short paths between any pair of components, highly clustered neighbourhoods and broad-tailed connectivity distributions \cite{Newman.10, Albert.Barabasi.02, Newman.rev03, Boccaletti.etal.06}. These properties are far from random and therefore indicate possible involvement of certain design constraints in the overall structures. Beyond the global features, it is possible also to identify local structural elements, the fundamental building blocks, which reflect the underlying process of network generation. We focus our study on the structural analysis of the networks of paired interactions in \textit{R. marginata}, specifically the dominance networks, and ask the following questions: how are the networks built? What is the underlying structural design? What is the function for which the networks are designed? What are the basic functional elements present in the networks? In particular, we search the networks for `motifs' \cite{Milo.etal.02}, basic units of interconnections that occur at frequencies significantly higher than those in their randomized counterparts. Earlier studies have demonstrated that information transfer networks such as gene transcriptional networks, neuronal connectivity networks and electronic circuit networks share common significant substructural patterns \cite{Milo.etal.02, Milo.etal.04}. In a recent study on seed harvester ant \textit{Pogonomyrmex californicus}, a social insect species, Waters and Fewell found the predominance of similar patterns in their antennation networks \cite{Waters.Fewell.12}. Our hypothesis about the function of the dominance networks of \textit{R. marginata} prompt us to expect similar patterns. 

\subsection{Boolean modeling}
Similarity in structural patterns of networks, nevertheless, does not guarantee a similar functionality \cite{Ingram.etal.06}. It would be necessary to investigate the dynamics of information flow in the networks and check whether the networks are robust and capable of efficient transfer of information. Boolean networks provide a useful framework to study generic dynamical systems with unknown or partially known structure or function \cite{Aldana.etal.02, Gershenson.04, Saadatpour.Albert.13}. The modeling scheme was first introduced by Stuart Kauffman for studying gene transcriptional networks \cite{Kauffman.69, Kauffman.93} and subsequently applied successfully in different organization levels of biological regulatory networks within organisms \cite{Kauffman.etal.03, Bornholdt.08, Bhardwaj.etal.11, Patarn.Carnev.87}, and also in various social networks \cite{Kluver.Scmidt.99, Green.etal.07}. We use Boolean modeling to analyze the dynamical behaviour of the networks and ask: how sensitive are the networks with respect to the smallest of perturbations? And how do the non-random architecture of the networks affect the efficiency of information transfer? These features are investigated by monitoring how the `Hamming distances', differences in information between a pair of predetermined initial states, evolve in time. Our investigation suggests that there is indeed a design principle involved in the dominance networks of \textit{R. marginata} that favours efficient transfer of information. We also believe that our analyses can help to develop an understanding about the evolutionary process by which the biological regulatory systems have been optimized.

\section{The experimental data} 

We used the data from behavioural observations carried out on $9$ post-emergent colonies of \textit{R. marginata}. The colonies were of different sizes ranging from 14 to 59 adults. The individuals on the nests were uniquely marked with spots of Testors\textsuperscript{\textregistered} quick drying enamel paints prior to the observations. The major activity period of the day for \textit{R. marginata} is of 10 hours, between 8 AM to 6 PM, which was divided into four equal blocks of 2 hours and 30 minutes each. Each colony was observed for 5 hours in a day in two such alternate blocks, over two consecutive days, covering the entire activity period, thus yielding 10 hours of data per colony. Blocks consisted of sessions lasting 5 minutes, followed by a break of 1 minute between every session. Observation sessions were of two kinds, either `instantaneous scan' sessions or `all occurrences' sessions, and they were randomly intermingled in the ratio of 1:2. In the former case a snapshot of the behavioural state of each individual was recorded and in the latter every occurrence of a set of chosen behaviours by any individual was noted down \cite{Gadagkar.01}.

\section{The dominance networks} 

\begin{figure*}
\centerline{\epsfig{file=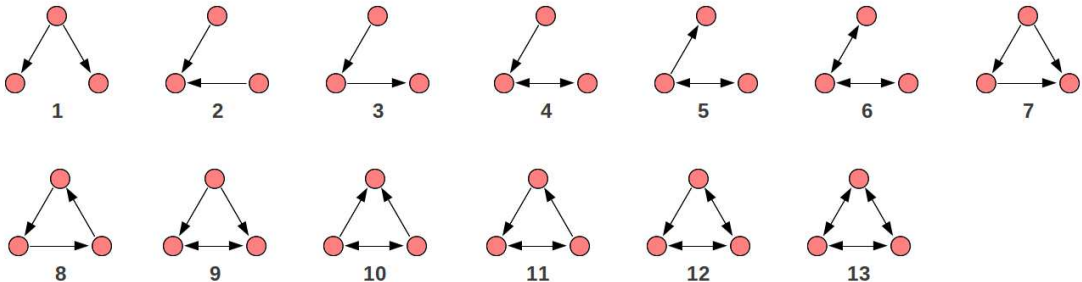,width=15cm}}
\caption{The $13$ possible connected three-node substructures (triads).} 
\end{figure*}

In the dominance networks of \textit{R. marginata}, all individuals were considered as nodes. A dominance relationship between a pair of individuals is represented by a directed link from the dominant to the recipient individual. Nine distinct behaviours shown by the individuals are termed as dominance behaviour in \textit{R. marginata}, they are: attack, chase, nibble, peck, crash land on another individual, sit on another individual, being offered regurgitated liquid, aggressively bite, and hold another individual in mouth \cite{Gadagkar.01}. Such dominance relationships, sampled by the 10 hours of observation, were used to construct the dominance networks for each of the $9$ colonies (Supplementary text 1). The network constructed for the colony V215 is shown in the figure 1. The basic quantities measured pertaining to the global structure of the networks are shown in the table I. The quantities include (i) \emph{average path length}: average number of links on the shortest path between any pair of nodes, (ii) \emph{clustering coefficient}: average density of links in the neighbourhood of the nodes, (iii) \emph{assortativity coefficient}: correlation among the connected nodes, and (iv) \emph{degree distributions}: probability distribution of a random node having specific connectivity for both \emph{in-degree} (the number of links that are directed towards a node) and \emph{out-degree} (the number of links that are directed away from the node).
  
In order to uncover the global structural patterns, each network was compared to a corresponding random network that preserved the number of nodes $N$ and links $L$ of the original network. Such random networks, known as Erd\H os-R\'enyi random graphs \cite{Erdos.Renyi.59}, are characterized by Poisson distributed in-degree and out-degree distributions with the ratio $L/N$ as the parameter. Such expected Poisson distribution was tested against the distributions of the original networks by means of $\chi^2$ goodness of fit test, and the corresponding probabilities are furnished in the table I. For both in-degree and out-degree, we found $P>0.1$ in $8$ out of $9$ colonies, which suggests lack of evidence in favour of the distributions to be different from Poisson. In a random network, the average path length is expected to grow logarithmically and the clustering coefficient is expected to fall inversely with the network size $N$, but the sizes of our colonies restricted the detection of such effects. The assortativity coefficient for random networks is expected to be zero, but the \textit{R. marginata} dominance networks show an average of $-0.23\pm0.15$ which is significantly less than zero ($t$-test, $P=0.001$). Therefore, the networks are said to be \emph{disassortative} \cite{Newman.03}, which means that the more dominant individuals tend to dominate individuals who are less dominant; in other words, highly dominant individuals tend to avoid other highly dominant individuals. It has been found that many biological networks engaged in information transfer such as protein interaction network in the yeast \textit{Saccharomyces cerevisiae}, metabolic network in the bacterium \textit{Escherichia coli}, neuronal connectivity network of the nematode worm \textit{Caenorhabditis elegans}, and technological networks like the Internet and the World-Wide-Web show dissasortative mixing in their network structure \cite{Newman.03}. It is possible that, in the course of information transfer, the components having information are more interested to pass it on to the functional components, rather than sharing it among themselves. Such a mechanism would certainly be economic and therefore favoured under selection mechanism. The analyses in this section were performed by using the network analysis software Cytoscape \cite{Smoot.etal.11}.
 
\section{Analysis of the substructures} 

\begin{table*}
\begin{tabular}[c]{|c|c|c|c|c|c|}
\hline 
\footnotesize{Triad} & \multicolumn{2}{|c|}{\footnotesize{Observed}} & \footnotesize{Observed} & \footnotesize{Significant} & \footnotesize{Less significant} \\ 
\footnotesize{identity} &\multicolumn{2}{|c|}{\footnotesize{frequency(\%)}} & \footnotesize{in networks} & \footnotesize{in networks} & \footnotesize{in networks} \\ \cline{2-3}
 & \multicolumn{1}{|c|}{\scriptsize{mean}} & \multicolumn{1}{|c|}{\scriptsize{sd}} &  & \scriptsize{(count $>1$, $P<0.1$)} & \scriptsize{(count $>1$, $P>0.9$)} \\
\hline 
1 & 36.66 & 9.35 & 9 & 0 & 4\\ 
\hline
2 & 21.72 & 10.47 & 9 & 1 & 4\\ 
\hline 
3 & 31.18 & 11.03 & 9 & 1 & 0\\ 
\hline 
4 & 0.00 & 0.00 & 0 & 0 & 0\\ 
\hline 
5 & 0.00 & 0.00 & 0 & 0 & 0\\ 
\hline 
6 & 0.00 & 0.00 & 0 & 0 & 0\\ 
\hline 
7 & 10.22 & 5.80 & 8 & 6 & 0\\ 
\hline 
8 & 0.22 & 0.49 & 2 & 0 & 0\\ 
\hline 
9 & 0.00 & 0.00 & 0 & 0 & 0\\ 
\hline 
10 & 0.00 & 0.00 & 0 & 0 & 0\\ 
\hline 
11 & 0.00 & 0.00 & 0 & 0 & 0\\ 
\hline 
12 & 0.00 & 0.00 & 0 & 0 & 0\\ 
\hline 
13 & 0.00 & 0.00 & 0 & 0 & 0\\ 
\hline 
\end{tabular} 
\caption{Frequencies of the triads and their significance in all the $9$ dominance networks of \textit{R. marginata} put together.}
\end{table*}

The three-node substructures, or the \emph{triads}, can be thought of as the basic building blocks of a network \cite{Wasserman.Faust.94}. The $64$ possible types of triads can be classified into $16$ isomorphic classes \cite{Wasserman.Faust.94}, out of which we are interested in those $13$ where all three nodes are connected (figure 2). Each network was searched for all of these $13$ triads and the number of occurrences of each triad was recorded. The measured quantities were then compared with those of a properly randomized network. This time, the randomized networks were constructed by keeping the single-node characteristics preserved such that both the in-degree and out-degree of each node remained unaltered. The measurement over the ensemble provides dispersions in the measured quantities and therefore allows statistical comparisons with the original \cite{Maslov.Sneppen.02, Maslov.etal.04}. The statistical significance was tested by using two different methods. First with the empirical sample estimate of probability $P$, which is defined as the probability that the particular triad appears in the randomized networks an equal or greater number of times than in the original network. The substructure is said to be significantly overrepresented in the real network and subsequently called a `motif', if $P$ is lower than a small predetermined cut-off value \cite{Milo.etal.02}. The under-representation of a triad also can be inferred if $P$ is found to be higher than some high cut-off. The other way to determine the statistical significance is to compute the \emph{Z-score}; the normalized deviation of the occurrences from the expected mean value. If $N_\mathrm{o}$ is the number of times the $i$th triad occurs in the original network, and $\bar{N}_\mathrm{r}$ and $\sigma_\mathrm{r}$ are the mean and standard deviation of its occurrence in the randomized networks, the the Z-score for the triad is defined as $Z_i=(N_\mathrm{o}-\bar{N}_\mathrm{r})/\sigma_\mathrm{r}$. A higher value of the Z-score of a triad implies a higher significance of the occurrence of that triad in the network \cite{Milo.etal.04}. Both the methods of testing statistical significance have their own limitations; distributions may be undersampled or may be different from a Gaussian \cite{Ziv.etal.05}, so we decided to use both the methods for our purpose. The analyses were done by using an application called the Fast Network Motif Detection (FANMOD) \cite{FANMOD.06}. The application made use of an algorithm named RAND-ESU which is an efficient algorithm especially in the case of substructures with low concentrations \cite{Wernicke.06, Nejad.etal.12}. The results are summarized in the table II. 

\subsection{The feed-forward loop}
\begin{figure*}
\centerline{\epsfig{file=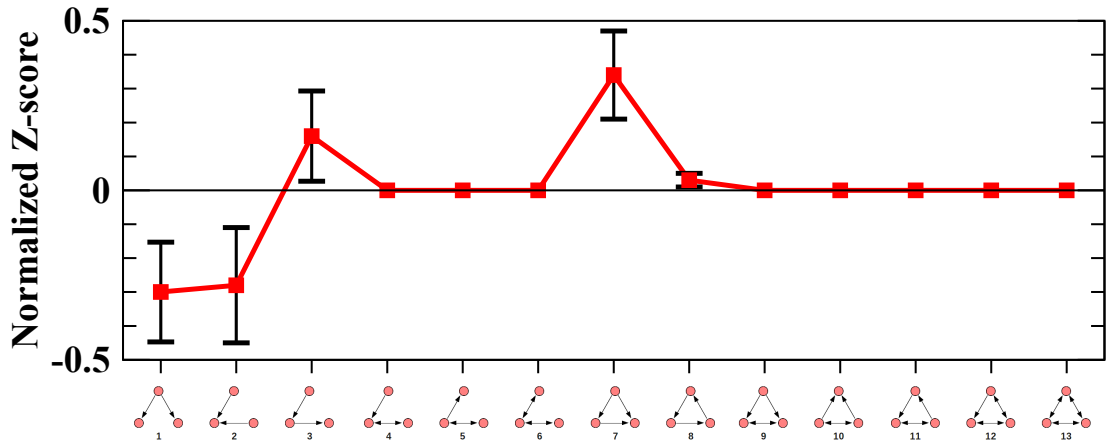,width=17cm}}
\caption{The `triad significance profile' for the dominance networks of \textit {R. marginata}. The normalized Z-scores corresponding to each kind of triad is averaged over $9$ colonies, whiskers show their standard errors of measurements, and the line connecting the Z-scores serves as a guide to the eye.} 
\end{figure*}

In the dominance networks of \textit{R. marginata}, the prominently found substructures were triad $1$, $2$, $3$ and $7$. All these triads were previously shown as the most structurally stable triads among the all $13$ possibilities \cite{Prill.etal.05}. The only other triad found, though very rarely, was triad $8$, which is one of the least stable triads \cite{Prill.etal.05}. It is worth noting that the number of occurrences of triad $7$ was much less compared to the other three predominating triads, but when compared with randomized networks, triad $7$ was found to be significantly overrepresented in $6$ networks (V213, V215, V219, V220, V223 and V224), while the significance of the others were negligible. Triads $1$ and $2$ were significantly underrepresented in $4$ colonies each. The triad $7$ showed averaged normalized Z-score $0.34\pm0.39$ which was significantly greater than zero ($t$-test, $P=0.015$), no other triads showed significant deviation from zero. By both the methods of significance testing, the triad $7$, commonly known as the feed-forward loop, emerged as the most consistently significant substructure present in the dominance networks. We would like to mention here that we have also performed similar analyses with other networks of paired behaviours such as antennation, allogrooming, soliciting and food sharing, but no triad was found consistently significant over the colonies. The results were found to be qualitatively similar when the analysis was repeated on the same nests with 30 hours of behavioural data.

The feed-forward loop was previously found to be significant in transcriptional networks of bacteria like \textit{Bacillus subtilis} and \textit{E. coli}, and the yeast \textit{S. cerevisiae} \cite {Milo.etal.02, Milo.etal.04, Lee.etal.02, Shen.etal.02, Ishii.etal.01}. The motif was also found in neuronal connectivity network of the nematode \textit{C. elegans} \cite {Milo.etal.02}. Since all these networks carry information from sensory components to the functional units, it has been argued that the structural elements common to them may play a functional role in information processing \cite {Milo.etal.02}. It has been shown, both by theory and experiment, that the feed-forward loop performs signal processing tasks such as persistence detection, pulse generation and acceleration of transcription responses \cite{Shen.etal.02, Mang.etal.03, Mang.Alon.03}. The same signature found in the antennation network of the seed harvester ant \textit{P. californicus} while engaged in foraging activities also suggests a functional similarity and selection for efficiency of directional information flow \cite{Waters.Fewell.12}. 

\subsection{The triad significance profile}
The comparison of Z-scores across the colonies is possible only after the data is subjected to normalization since the size of the networks usually have substantial influence on their absolute values \cite{Milo.etal.02}. Therefore, for each colony, the vector of Z-scores was normalized to a length unity by computing $Z_i/\sqrt{\Sigma_i Z_i^2}$ for each of the triad. These normalized Z-scores for each triad were averaged over all the colonies to get the `triad significance profile' of the species \cite{Milo.etal.04}; the profile is shown in the figure 3. Except for triad $3$, the profile shows fair resemblance with the superfamily of sensory transcriptional networks that controls gene expression in bacteria and yeast in response to external stimuli \cite{Milo.etal.04}. This similarity suggests that the networks may have evolved under similar constraints to perform tasks in a similar manner \cite{Milo.etal.04}. The sensory transcriptional networks are rate-limited networks, where the response time of each step in the networks are of the order of the response time required for the functioning of the networks \cite{Milo.etal.04}. If the dominance networks of \textit {R. marginata} are used for the transfer of the colony level information and subsequently for the regulation of worker activity, the networks are indeed rate-limited, since the response time for the workers to a particular task is expected to be as short as the response time of each interaction. It should be worth noting that the profile is largely different from the other superfamily of biological information processing networks that are not rate-limited and also from the superfamily of social networks \cite{Milo.etal.04}. Both these superfamilies are characterized by triads containing mutual links which do not appear in the our networks. Therefore, we would like to infer that the specific non-random structure of the dominance networks of \textit{R. marginata} have evolved under selection mechanisms similar to the other biological rate-limited regulatory networks.

\section{Analysis of the dynamics}  
\begin{figure*}
\centerline{\epsfig{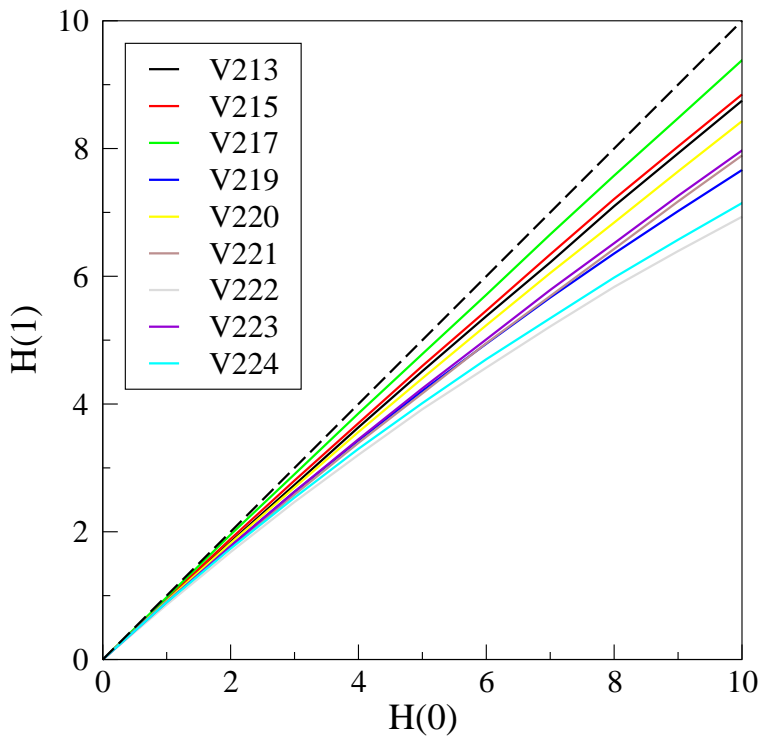}}
\caption{Evolution of initial Hamming distances, $H(0)$ with one time step to $H(1)$, for each of the $9$ networks. Each network dynamics is averaged over $10^5$ different initial random configurations. The dashed line signifies $H(1)=H(0)$. Legends show colony identities (Online version in colour).} 
\end{figure*}

In a Boolean network model, each node can be in any of the two possible states, `active' and `inactive', represented by the binary values $1$ and $0$ respectively. In the dominance networks of \textit{R. marginata}, the nodes with some information are considered as active and the nodes with no information are considered as inactive. The information about the colony needs may not necessarily be a binary variable, but this simplification is justified since a node can become active only when a certain threshold value of the information is reached, and remain inactive till the information is below the threshold \cite{Aldana.etal.02}. Such threshold models for worker activity have already been tested in social insect species \cite{Bona.etal.96, Bona.etal.98}. We assumed that the information can pass only from the dominant to the subordinate individual, therefore the future state of a node is regulated by those nodes to which it is connected with incoming links. As the system evolves with time, the state of a node is updated according to a coupling between the node and the nodes dominating over it. These couplings are Boolean functions expressed generally through a combination of the logic operators AND, OR, and NOT \cite{Aldana.etal.02, Saadatpour.Albert.13}. In a classical Boolean network model, the interconnections and the interaction rules are probabilistically chosen from predefined sets. However, having some understanding of the system, it is customary to use that knowledge to choose the rule that best describes the system \cite{Saadatpour.Albert.13, Wang.etal.12}. We assumed that a single dominance interaction is sufficient to pass the information from the dominant to the subordinate individual, therefore we used the most generic form of OR logic as the interaction rule, hoping that the real systems share their most important properties with the most generic representation we are dealing with \cite{Aldana.etal.02}. According to the OR rule, if any one of the input node is active, the output becomes active, otherwise it is inactive; and a node with no incoming connection retains its original state. Starting from a random configuration of active and inactive nodes, the system keeps on updating following the interaction rules until the system reaches a steady state. The synchronous updating scheme used is the simplest and most convenient scheme, where at each time step, the states of all nodes are updated simultaneously depending on the states at the previous time step. The fact that the durations of the dominance interactions are mostly alike supports such an updating rule \cite{Kauffman.etal.03, Wang.etal.12}.
       
\subsection{The Hamming distances}
To study the stability of the system in terms of information transfer, we investigated the time development of small fluctuations in the system. We started with a pair of different possible initial states $\Sigma_0=\{\sigma_1(0),\sigma_2(0),...,\sigma_N(0)\}$ and $\tilde{\Sigma}_0=\{\tilde{\sigma}_1(0),\tilde{\sigma}_2(0),...,\tilde{\sigma}_N(0)\}$ sampled from the entire state space, where $\sigma$'s are the binary values of the states ($0$ or $1$) for each of the $N$ nodes at $t=0$. The Hamming distance at time $t$, defined as $H(t)=\sum_{i=1}^N (\sigma_i(t)-\tilde{\sigma}_i(t))^2$, was then plotted against predefined initial distances $H(0)$; the representation scheme is sometimes referred in the literature as Derrida plot \cite{Kauffman.etal.03, Derrida.Weisbuch.86}. The slope of the curve near the low $H$ region reflects the fate of small fluctuations in the system. If the curve is well above the $H(t)=H(0)$ line, then the system transfers information to a number of nodes that grow exponentially with time and the system is said to be in a \emph{chaotic} phase, which is an unstable state and easily prone to noise. On the other hand, if the curve is well below the line, the fluctuation decays exponentially and therefore propagates to only a few number of other nodes and the system is said to be in a frozen or \emph{ordered} phase. In the intermediate situation, very near the line, the information flows to a number of nodes that grow algebraically with time and the system is said to be in the edge of chaos or in the \emph{critical} phase. For efficient transfer of information, the system is expected to be in an ordered phase near the edge of chaos \cite{Gershenson.04}. The figure 4 shows the plot for each of the $9$ colonies, with $t=1$. All the curves are below but near the $H(1)=H(0)$ line, suggesting the stability against small perturbations and efficient flow of information.
  
\begin{figure*}
\centerline{\epsfig{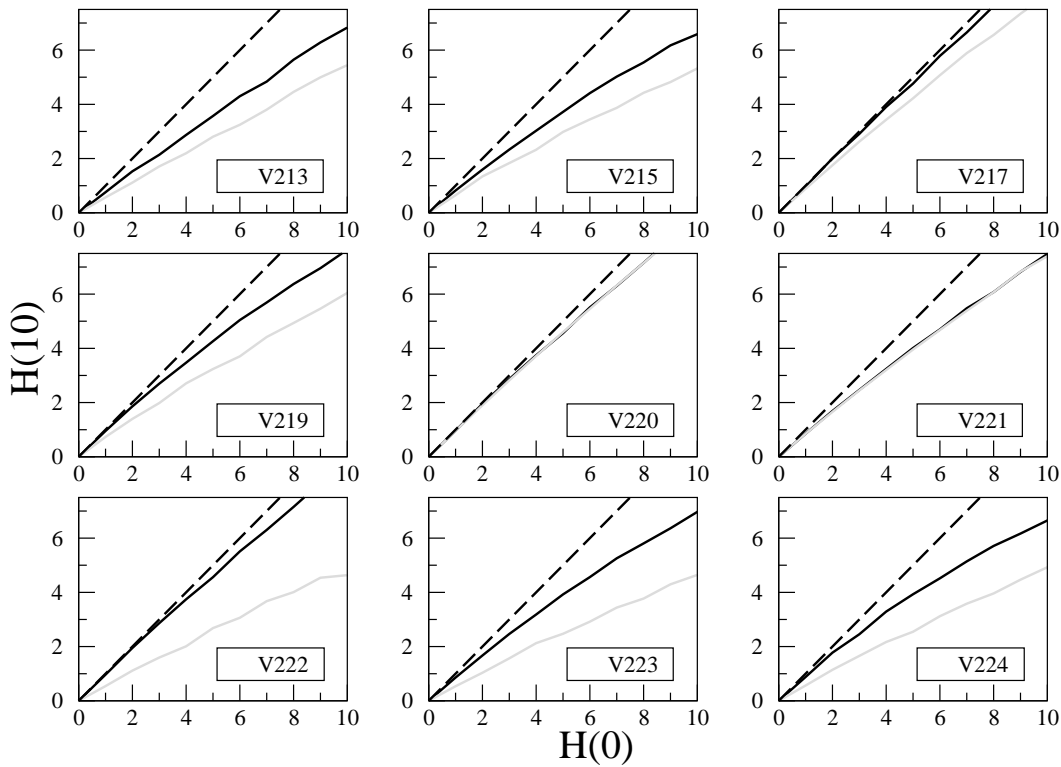}}
\caption{Evolution of initial Hamming distances, $H(0)$ with $10$ time steps to $H(10)$, for each of the $9$ networks. Black solid curves represent original networks, each averaged over $10^4$ random initial conditions. Each original network is randomized $10^2$ times and each of them is averaged over $10^4$ initial configurations (gray solid curves). The dashed lines signify $H(10)=H(0)$. Legends show colony identities.} 
\end{figure*}

\subsection{The effect of design}
To examine the influence of network architecture over the information transfer dynamics, the time evolution of the Hamming distance as a function of initial distance was compared to that of the randomized version \cite{Maslov.Sneppen.02} of the network. If the architecture of the original networks influence information transfer, we expect to get patterns different from those of the randomized networks \cite{Kauffman.etal.03}. The differences, if any, would be prominent only after $t=1$ since the changed outputs could be used as inputs only after the first time step \cite{Kauffman.etal.03}. We chose an arbitrary duration $t=10$ and plotted $H(10)$ against $H(0)$ for both the original and randomized networks in the figure 5. We found that the original networks (black curves) are near the critical phase, while the randomized counterparts (gray curves) are generally in more ordered phase. In particular, the randomized curves are significantly below the original curves in $7$ colonies (V213, V215, V217, V219, V222, V223 and V224, $t$-test, all $P<0.0005$) while the other $2$ are indistinguishable ($t$-test, both $P>0.4$).

We also investigated changes in predetermined Hamming distances $H(t)$ over time (figure 6). Starting from $H(t)=1$ at $t=0$, we plot the mean Hamming distances $H(t)$ as function of time $t$, for both the original networks (black curves) and the randomized networks (gray curves). We found that in the same $7$ colonies, the randomized $H(t)$ is significantly below the original $H(t)$ ($t$-test, all $P<0.0005$). In the results depicted in the figures 5 and 6, standard errors of measurements were found to be of the order of $0.01$ or less, so they are not visible in the figures. Therefore, at least in $7$ colonies, the original network is less-frozen than its randomized counterpart, which means that in the randomized networks information dies out more quickly and the original network is comparatively more capable of holding information. Since the structural pattern of the networks get destroyed in the randomization process, this less-frozen property of the original networks could be attributed to their distinctive structure. This attribution is well supported by the observation that, out of the $7$ less-frozen colonies, $5$ are in common with the $6$ colonies where the feed-forward loop was observed significantly more than random. The mean normalized Z-score of the feed-forward loop for the $7$ less-frozen colonies is $0.43\pm0.26$ which is larger than that for all the $9$ colonies taken together. Therefore, we would like to infer that the dominance networks of \textit{R. marginata} are designed for the specific purpose of information transfer with the `feed-forward loop' being the key functional element.

\section{Conclusions and future directions}
\begin{figure*}
\centerline{\epsfig{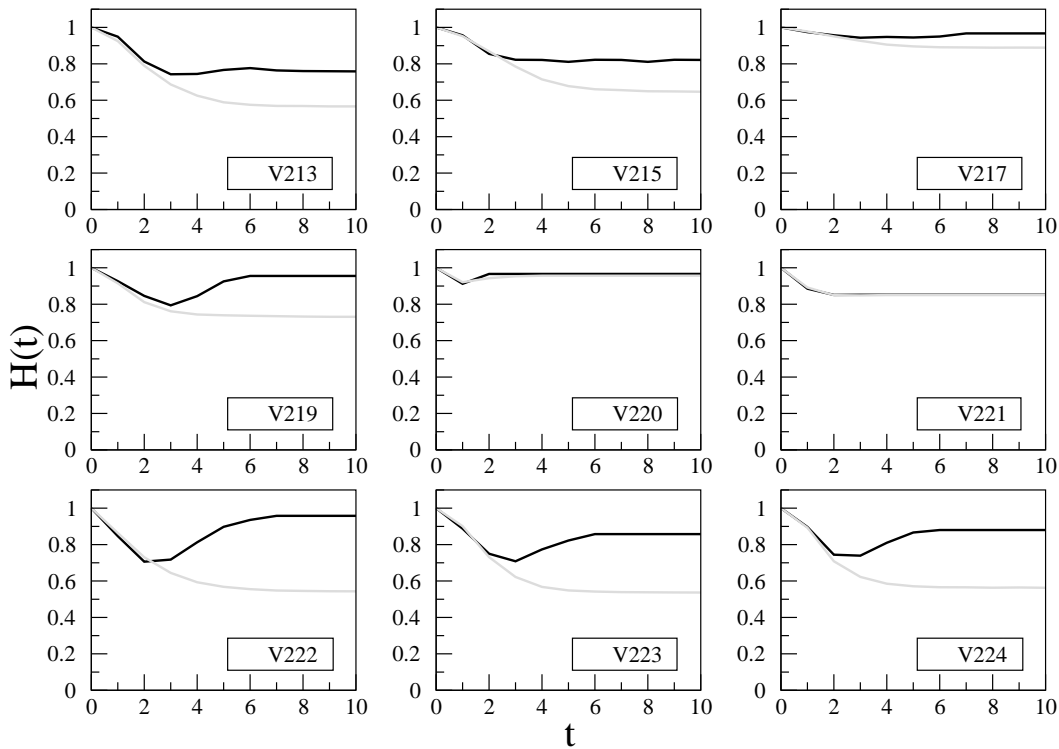}}
\caption{The time evolution of the Hamming distances starting from $H(0)=1$. Black curves represent original networks, each averaged over $10^4$ random initial configurations. Each original network is randomized $10^2$ times and each of them is averaged over $10^4$ initial configurations (gray curves). Legends show colony identities.} 
\end{figure*}

In recent years, network substructure analysis of complex systems has gained much attention among the scientists. Several biological and technological networks have been grouped into superfamilies based on similarities in the statistics of possible substructures. It has been noticed that the members of the same superfamily perform similar tasks. Our investigation into the local structure of the networks constructed from the dominance interactions in \textit{R. marginata} allows to draw parallels with a superfamily of naturally evolved information processing networks. In the latter networks and in the dominance networks of \textit{R. marginata} as well, the feed-forward loop appears to be statistically overrepresented. In contrast, the networks constructed from paired interactions in \textit{R. marginata} other than dominance fail to show any significant triad. This result is a supporting evidence favouring the role played by dominance interactions in information processing. However, the presence of substructures alone can not ensure the overall functioning of a network. To understand whether the information transfer through dominance can indeed be called efficient, we model the dynamics of information transfer using simple Boolean functions and compare our results against a suitably randomized ensemble. Our result supports the idea that there exists common evolutionary design principles by which the biological regulatory networks are optimized. Further research along this direction would allow for more accurate prediction of the properties of a newly identified network on the basis of other networks in the same superfamily. 

On the other hand, we have tried to reason the existence of the dominance behaviour, the purpose of which still remains an intriguing aspect among the evolutionary biologists in the context of colony organization \cite{Izzo.etal.10, Dappo.etal.10}. The queen in the most studied primitively eusocial wasp \textit{Polistes dominulus} is known to use dominance to gain reproductive monopoly over her colony members \cite{Pardi.48}. But in the colonies of tropical species \textit{R. marginata}, the queen usually shows very little or no dominance at all \cite{Sumana.Gadagkar.03}, though a clear dominance hierarchy can be recognized in the colony \cite{Gadagkar.80}. Such dominance behaviour among the workers cannot be associated with the reproductive competition because there is no correlation between the dominance rank of an individual and the probability that she will replace a lost or removed queen \cite{Chandra.Gadagkar.92}. With some experimental correlates, it has been argued earlier, that the workers use dominance to transfer information about the colony needs to their co-workers, which is essential for their self-organized regulation of work. However, there has been no previous study of how efficient such a mechanism can be. This study shows that the dominance networks of \textit{R. marginata} are indeed designed for efficient information transfer, and hence might be used for self regulation, not only for meeting the colony demands for food, but also for other maintenance purposes like nest building, thermoregulation, defense etc. However, we should mention that a more direct approach would be a sequential analysis of dominance behaviour and the related worker activities. Therefore, a plausible hypothesis that could be favoured by our findings is that, in the course of evolution, the dominance behaviour in the social wasps has been adapted for information transfer in addition to its primal function of reproductive monopoly maintenance. This hypothesis would needed to be tested in other social insects species as well.

\section{Acknowledgments}

This study was supported by the Department of Biotechnology, Department of Science and Technology, Ministry of Environment and Forests, Council for Scientific and Industrial Research, Government of India, and BITS Pilani Research Initiation Grant Fund. The data analyses, modeling, simulations and statistical tests were performed by AKN and KB. Behavioural observations on \textit{Rm} were carried out by AS. The paper was co-written by AKN and KB. Authors wish to thank Prof. Raghavendra Gadagkar, CES, IISc, Bangalore, India for his valuable support and guidance throughout the entire work. Authors also wish to thank Dr. Jennifer H. Fewell, ASU, Tempe, USA for initial discussion regarding this problem and also to Prof. Tam\'as Vicsek, EU, Budapest, Hungary, Dr. Anindita Bhadra and Dr. Ayan Bannerjee, IISER, Kolkata, India, and Miss. Paromita Saha, IISc, Bangalore, India for their generous help to improve the manuscript. All experiments reported here comply with the current laws of the country in which they were performed. All statistical tests were done by using the statistical environment R \cite{R.13}.

\end{document}